\documentclass[10pt,a4paper,twocolumn]{article}

\usepackage{newtxtext}   % Times text
\usepackage{newtxmath}   % matching Times math
\usepackage[T1]{fontenc}
\usepackage[utf8]{inputenc}

\usepackage[a4paper,top=2.5cm,bottom=2cm,left=2.5cm,right=2.5cm]{geometry}

\usepackage{graphicx}
\usepackage{float}      % [h] specifier: pin figures exactly where written
\usepackage{amsmath}
\usepackage{booktabs}
\usepackage{siunitx}
\usepackage{textcomp}
\usepackage{tikz}
\usepackage{xcolor}
\usepackage[hidelinks]{hyperref}
\usetikzlibrary{arrows.meta,calc,positioning,decorations.pathreplacing}

\definecolor{g14gas}{RGB}{232,140,60}     % plasma / high-Mach gas
\definecolor{g14solid}{RGB}{70,130,180}  % solids / heat conduction
\definecolor{g14part}{RGB}{70,160,90}    % Lagrangian particles
\definecolor{g14rad}{RGB}{196,160,40}    % volumetric radiation

\usepackage{titlesec}
\titleformat{\section}{\normalfont\fontsize{11}{13}\selectfont\bfseries}{\thesection}{0.6em}{}
\titleformat{\subsection}{\normalfont\fontsize{10}{12}\selectfont\bfseries}{\thesubsection}{0.6em}{}
\titleformat{\paragraph}[runin]{\normalfont\fontsize{10}{12}\selectfont\bfseries}{}{0em}{}
\titlespacing*{\section}{0pt}{6pt}{2pt}
\titlespacing*{\subsection}{0pt}{4pt}{1pt}
\titlespacing*{\paragraph}{0pt}{4pt}{0.6em}

\usepackage[font={small,it},labelsep=period,justification=centering]{caption}
\usepackage{enumitem}
\usepackage{etoolbox}
\apptocmd{\thebibliography}{\setlength{\itemsep}{0pt}\setlength{\parsep}{0pt}\setlength{\parskip}{0pt}}{}{}

\begin{document}

% ====================================================================
%  FULL-WIDTH TITLE BLOCK  (spans both columns)
% ====================================================================
\twocolumn[
  \begin{center}
    {\fontsize{16}{19}\selectfont\bfseries
     Digital Twin of an Argon--Hydrogen Plasma Reactor\par}
    \vspace{6pt}
    {\fontsize{11}{13}\selectfont
     B.~Artola\textsuperscript{1}, A.~Bonfanti\textsuperscript{1}, 
     M.~Holmstr\"om\textsuperscript{1},
     Q.M.~Wargnier\textsuperscript{1}\par}
    \vspace{3pt}
    {\fontsize{10}{12}\selectfont
     \textsuperscript{1}GREEN14 AB,
     Brinellv\"agen 25, SE-114 28 Stockholm, Sweden.\par}
  \end{center}
  \vspace{6pt}

  % --- Abstract + keywords: full page width (spans both columns) ---
  {\fontsize{11}{13}\selectfont\bfseries Abstract\par}
  \vspace{2pt}
  The principal proof of concept revolves around an argon--hydrogen plasma reactor that melts, reduces, atomizes and quenches critical raw material in one step, with premium spherical powder as the deliverable output and control of the composition chemistry. Each usage of the reactor is monitored through thermocouples and pressure sensors, which provide a daily data source of the real-world experiments. The reactor is modeled through COMSOL Multiphysics\textsuperscript{\textregistered}, which represents the core solver used to provide multiphysics simulations. The usage of COMSOL is complemented with Artificial Intelligence (AI) models, to enable seamless data assimilation and optimization. This paper presents the COMSOL twin of the reaction chamber and converging--diverging nozzle, together with a custom phase-change particle-tracing layer validated on Ti-6Al-4V (Ti64). Moreover, we highlight how the synergy between COMSOL simulations and AI-based digital surrogates can be leveraged to build self-consistent optimization loops geared toward (i)~fully autonomous live control of the reactor and (ii)~optimization of the process.\par

  \vspace{4pt}
  {\fontsize{11}{13}\selectfont\bfseries Keywords:\ }%
  {\fontsize{10}{12}\selectfont
  FEM, RANS, Argon--Hydrogen Plasma, Particle Tracing, Phase Change,
  Ti-6Al-4V, Conjugate Heat Transfer, Supersonic Nozzle, Digital Twin.\par}
  \vspace{6pt}
]

% ====================================================================
%  INTRODUCTION
% ====================================================================
\section{Introduction}
GREEN14\footnote{\url{https://www.green14.com}} develops an argon--hydrogen plasma reactor that melts, reduces, atomizes and quenches critical raw material in one continuous step, delivering spherical powder with controlled composition chemistry. This work is built around that reactor: an arc-heated Ar/H$_2$ jet that processes injected powder in a water-cooled chamber and expands through a converging--diverging nozzle. Plasma-torch spheroidization and atomization are established routes to spherical metal and ceramic powders for additive manufacturing~\cite{hao2021rfti64,liu2026plasma}, in which nozzle clogging from powder deposition on cold walls is a recurring failure mode~\cite{qiu2023triple}.

The reactor is a tightly coupled multi-scale, multi-physics system: continuum gas dynamics, conjugate heat transfer, radiation and Lagrangian phase change coexist, from the millimeter-scale nozzle to the chamber and from the Eulerian jet to the particles. High core temperatures prevent in-situ diagnostics at the reaction zone, so sensors sit far from the core and a multiphysics simulation is essential to reconstruct each experiment.

We model the reactor with the FEM implementation of COMSOL Multiphysics\textsuperscript{\textregistered}~\cite{comsol2024cfd} and compare against temperatures and pressures measured on the rig. Parameters that cannot be probed at the gun exit, notably the delivered inlet power, are inferred by matching the simulation to the real-world data, using Automatic Differentiation (AD) in JAX~\cite{sapunov2024deep} for exact sensitivities. The calibrated simulations then feed AI models that (i)~infer uncertain PDE parameters from real-world data, (ii)~recover the analytic form of missing source terms, and (iii)~keep the twin consistent with the rig. The core remains the physics model, not a black-box trained on a simulation dataset. A fast neural-operator layer~\cite{kovachki2023neural} can wrap that identified model for live control. This approach applies to other multi-scale, multi-physics problems, including other plasma processes.

This work presents a COMSOL model of the GREEN14 argon--hydrogen plasma reactor for Ti64 in the hot-plasma window, where the working fluid is a \emph{single} Ar/H$_2$ mixture. The production target is spherical Ti64, typically $15$--$45\,\mu\mathrm{m}$ (LPBF). Results focus on the chamber, the converging--diverging nozzle, and a custom phase-change particle-tracing layer that fills a gap in the \textit{Particle Tracing for Fluid Flow Module}~\cite{comsol2024particle}. The DC plasma gun, modeled separately with multifluid MHD, enters as a supersonic inlet; collection is omitted. The remainder of the paper is organized as follows. Section~\ref{sec:setup} describes the reactor and the modeling assumptions; Section~\ref{sec:methods} presents the governing equations and Section~\ref{sec:numerics} the discretization. Section~\ref{sec:AI} then identifies the uncertain coefficients against the experiments. Results follow in Section~\ref{sec:results} and are discussed in Section~\ref{sec:discussion}.
%(no solid--liquid--vapor phase change out of the box).

% ====================================================================
%  THEORY / EXPERIMENTAL SET UP
% ====================================================================
\section{Reactor configuration and assumptions}
\label{sec:setup}

\begin{figure}[H]
  \centering
  \includegraphics[width=\linewidth]{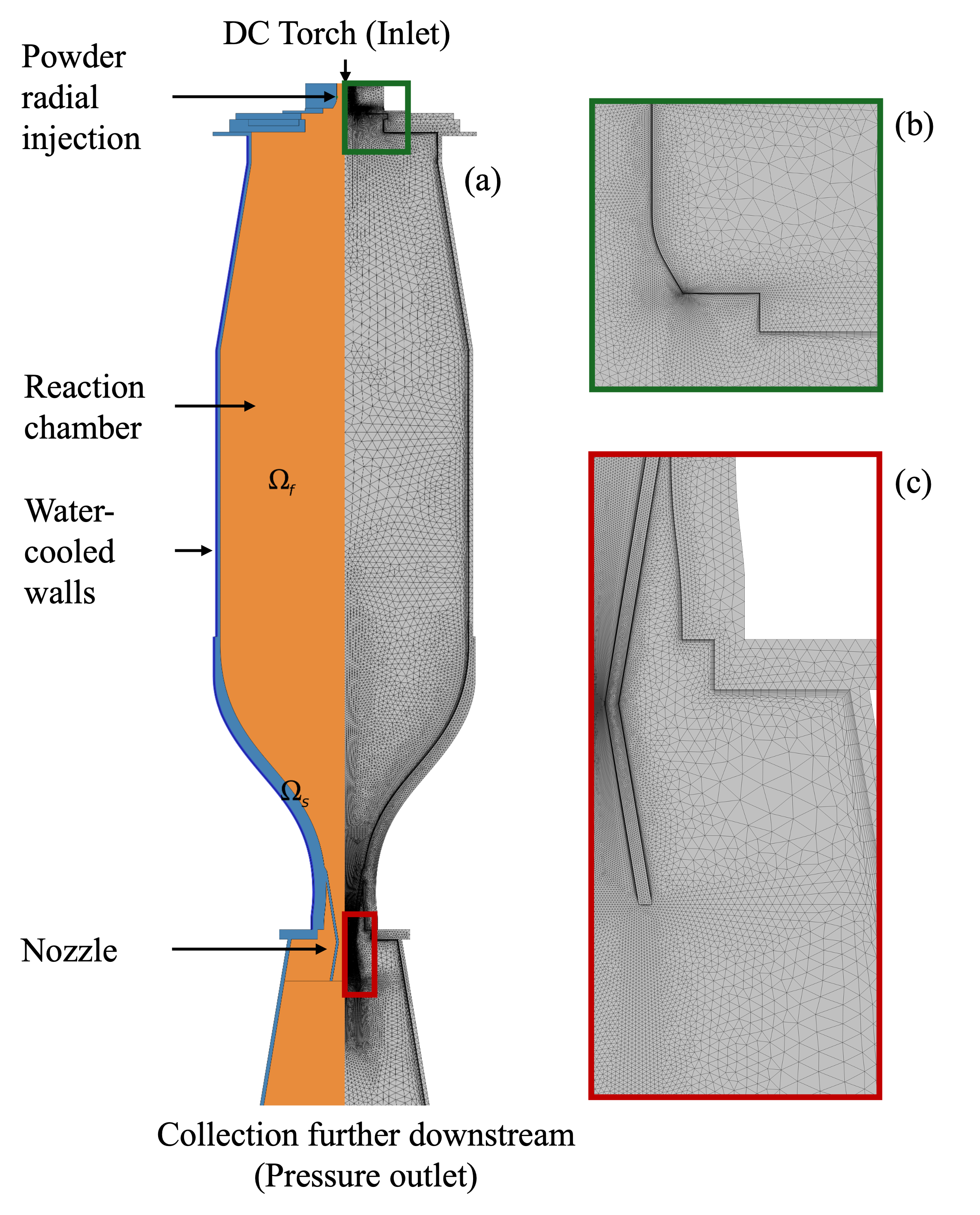}
  \caption{GREEN14 argon--hydrogen plasma reactor. (a)~axisymmetric half-section: fluid domain
  $\Omega_f$ and solid walls $\Omega_s$ on the left, computational mesh on the right.
  (b)~zoom on the inlet region (green box). (c)~zoom on the nozzle throat (red box).}
  \label{fig:reactor}
\end{figure}

Figure~\ref{fig:reactor} shows the GREEN14 argon--hydrogen plasma reactor and its computational mesh. From top to
bottom: (1)~DC plasma gun; (2)~reaction chamber, where feedstock is injected, melts and begins to evaporate;
(3--5)~converging--diverging nozzle that chokes, accelerates the gas and quenches the flow;
(6)~powder collection. This paper models (2)--(5) only. The gun is a supersonic inlet
(Section~\ref{sec:bc}); collection is omitted. Chamber and nozzle are resolved together: the
supersonic expansion quenches the powder, while the high slip in the throat breaks molten droplets.

The twin is exercised in the hot-plasma window: of order $10^2$\,NLPM of Ar with a few NLPM of
H$_2$, at tens of kilowatts; the chamber pressure is of order $500$\,mbar and the modeled
outlet is held at $75$\,mbar. Core temperatures drop
from $\sim$10$^4$\,K at the gun exit to a few $10^3$\,K in the chamber. The jet is sonic in the
throat and supersonic in the divergent section (Ma~$\approx 2$--$3$, diamond shocks), with
jet speeds $10^2$--$10^3$\,\si{m.s^{-1}} and $\mathrm{Re}\sim 10^5$--$10^6$.

% ====================================================================
%  NUMERICAL MODEL / METHODS / USE OF SIMULATION APPS
% ====================================================================
\subsection{Assumptions and gas properties}
\label{sec:assumptions}
The computational domain is the axisymmetric section
$\Omega=\Omega_f\cup\Omega_s\subset\mathbb{R}^{2}$ (Figure~\ref{fig:reactor}):
$\Omega_f$ is the plasma and $\Omega_s$ the water-cooled solid walls. Radiation enters as a
volumetric source in $\Omega_f$ (Section~\ref{sec:flow}). The twin retains three physics on
this split: compressible turbulent flow in $\Omega_f$, heat conduction in $\Omega_s$, and
Lagrangian phase-change particles. The DC gun is modeled separately with multifluid MHD in
the \textit{Plasma Module}; here it is replaced by a single-fluid supersonic inlet on which
the Ar/H$_2$ mixture enters as one LTE fluid (Section~\ref{sec:bc}).

Coupling is \emph{weak}: flow and solid conduction exchange temperature and heat flux at
$\Gamma_{\mathrm{fs}}=\partial\Omega_f\cap\partial\Omega_s$, but particles see the Eulerian
fields one way (no back-coupling of drag, heat or vapor). This is justified by dilute
loading: the injected powder mass, and its bulk density in the chamber, is much smaller
than that of the carrier gas, so the particles are passively advected and do not load the
mean flow.

The gas is a high-temperature LTE Ar/H$_2$ mixture: a prescribed argon mole fraction
$x_{\mathrm{Ar}}$, with equilibrium composition and mixture properties
($\lambda$, $\mu$, $C_p$, $M_n$, $H$) tabulated from Cantera~\cite{cantera} in
$(x_{\mathrm{Ar}}, T, p)$ over $300$--$30\,000$\,K. Species mass fractions come from those
tables rather than as PDE unknowns. Dissociation and ionization make $M_n$
(hence $R_s=R/M_n$, with $R$ the universal gas constant) strongly $T$-dependent. The tables
have been validated against previous plasma-physics approaches, including solar
physics~\cite{wargnier2020transport,wargnier2022collision}. Energy enters the domain as
inlet enthalpy (Section~\ref{sec:bc}). Joule heating, ionized-gas electrical conductivity
and the arc are left to the gun: the aim is a representation of the physical system for
process-engineer support and scale-up. Turbulence is closed with RANS
(Section~\ref{sec:flow}).

% ====================================================================
%  MODELING AND GOVERNING EQUATIONS
% ====================================================================
\section{Modeling and governing equations}
\label{sec:methods}

This section presents the PDEs solved in COMSOL for the gas, the particles and the solid walls.
The closures are the standard, robust COMSOL defaults for this class of compressible
conjugate problems, chosen because they are sufficient for a representative model of the
physical system that process engineers can use for support and scale-up.

\subsection{Fluid flow dynamics}
\label{sec:flow}
The goal is a mean-flow representation of the plasma in $\Omega_f$ that can be
compared with the rig measurements and used in an industrial calibration loop.
At $\mathrm{Re}\sim 10^5$--$10^6$ this is obtained by closing turbulence with
compressible RANS; coefficients that the closure leaves undetermined are
identified in Section~\ref{sec:AI}.

COMSOL's \emph{High Mach Number Flow, SST}~(\texttt{hmnf})
interface~\cite{comsol2024cfd} solves compressible mass, momentum and energy together with
Menter SST and an auxiliary wall-distance equation. The mean density $\rho$, velocity
$\mathbf u$ and temperature $T$ of the Ar/H$_2$ plasma satisfy
\begin{align}
  \partial_t\rho + \nabla\!\cdot(\rho\mathbf u) &= 0, \\[1pt]
  \rho\bigl(\partial_t\mathbf u + (\mathbf u\!\cdot\!\nabla)\mathbf u\bigr)
    &= -\nabla p + \nabla\!\cdot(\boldsymbol\tau + \boldsymbol\tau_R), \\[1pt]
  \rho C_p\bigl(\partial_t T + \mathbf u\!\cdot\!\nabla T\bigr)
    &= \nabla\!\cdot\bigl((\lambda + \lambda_T)\nabla T\bigr) \notag\\
    &\quad + \alpha_p T\bigl(\partial_t p + \mathbf u\!\cdot\!\nabla p\bigr)
     + \dot Q_{\mathrm{rad}},
\end{align}
closed by the ideal-gas law $p = \rho R_s T$ with $R_s = R/M_n$, where $p$ is the static
pressure. $C_p$ is the specific heat at constant pressure, $\alpha_p$ the isobaric expansion
coefficient, $M_n$ the mixture mean molar mass, and $\lambda$ and $\mu$ the molecular thermal
conductivity and dynamic viscosity, and $\mathbf I$ the identity. The Newtonian stress is
$\boldsymbol\tau = \mu\bigl(\nabla\mathbf u + (\nabla\mathbf u)^{\!\top}
-\tfrac23(\nabla\!\cdot\!\mathbf u)\mathbf I\bigr)$; $\boldsymbol\tau_R$ follows Boussinesq with
eddy viscosity $\mu_T$, and $\lambda_T = \mu_T C_p/\mathrm{Pr}_T$ (Kays--Crawford,
$\mathrm{Pr}_T\approx 0.85$). Pressure work and viscous heating are retained, as is the
mean-flow kinetic energy (required in the transonic/supersonic regime).

Radiation enters the energy equation as the volumetric source $\dot Q_{\mathrm{rad}}$,
closed with an optically thin net-emission-coefficient (NEC)
approximation~\cite{cressault2010nec},
\begin{equation}
  \dot Q_{\mathrm{rad}} = -\bigl[\,x_{\mathrm{Ar}}\,Q_{\mathrm{Ar}}(T)
  + (1-x_{\mathrm{Ar}})\,Q_{\mathrm{H_2}}(T)\,\bigr],
\end{equation}
a mole-fraction mix of tabulated emission coefficients that already include local
self-absorption.

Turbulence is closed with Menter's $k$--$\omega$ SST
model~\cite{Menter1994SST,Menter2003SSTHeatTransfer}, which blends a near-wall $k$--$\omega$
branch (wall heat flux) with a free-shear $k$--$\varepsilon$ branch (the jet). We use
COMSOL's low-Re SST with a boundary-layer mesh built so that the first-cell wall resolution
in viscous units satisfies $\Delta_w^{+} = \rho\,u_{\tau}\,\delta_w/\mu < 1$, where
$\delta_w$ is the wall-normal size of the first cell and
$u_{\tau} = \sqrt{\tau_w/\rho}$ the friction velocity from the wall shear stress $\tau_w$
(Figure~\ref{fig:mach}). The SST
constants are the standard Menter set; the quantities inferred from the rig are the few
engineering coefficients of the energy boundary conditions ($P_{\mathrm{in}}$,
$h_{\mathrm{contact}}$), identified in Section~\ref{sec:discussion}.

\subsection{Particle phase change}
\label{sec:particles}
The \textit{Particle Tracing for Fluid Flow Module}~\cite{comsol2024particle} provides
liquid-to-vapor evaporation. We extend it so that
the particles undergo the full solid--liquid--vapor transition: melting is added ahead of the
built-in evaporation, the drag law is switched to a compressible, high-Mach-number closure, and
molten droplets break up in the supersonic nozzle. A star $(\cdot)^{\star}$ is a per-particle
Lagrangian quantity. Coupling is one-way: particles see $\mathbf u$, $T$ and $p$ with no
back-coupling, as justified in Section~\ref{sec:assumptions}. A particle crosses the reactor in
about $0.1$\,s; the settled population (every particle having either left through the outlet
or frozen on a wall) on this frozen background is taken as the effective steady state of the
injected phase.

Particles follow the measured feedstock PSD. Feedstock size and shape are characterized by
dynamic image analysis (Sympatec QICPIC, dry dispersion): the size is the area-equivalent
circle diameter, reported as a volume-weighted ($q_3$) distribution, and the same analysis
returns a projected-image sphericity for each particle. This measured sphericity is taken as
the initial value $S^{\star}_0\in(0,1]$ of the Wadell sphericity
$S^{\star}= \pi^{1/3}(6V^{\star})^{2/3}/A^{\star}$~\cite{wadell1935} that enters the
Haider--Levenspiel drag closure~\cite{haider1989drag}.
Once the particle has melted it is taken as a sphere and kept as one, even if it later
solidifies, through the highest melt fraction reached up to time $t$,
\begin{equation}
  S^{\star}(t) = S^{\star}_0 + (1-S^{\star}_0)\,
  \max_{0\le\tau\le t} X_{\mathrm{melted}}^{\star}(\tau).
\end{equation}

Each particle is a point with mass $m^{\star}$, velocity $\mathbf v^{\star}$ and
temperature $T^{\star}$. Neglecting particle--particle interactions,
\begin{align}
  m^{\star}\frac{d\mathbf v^{\star}}{dt} &= \mathbf F^{\star}, \\[1pt]
  m^{\star} C_p^{\star}\frac{dT^{\star}}{dt} &= Q_c^{\star} + Q_r^{\star} - Q_p^{\star}, \\[1pt]
  \frac{dm^{\star}}{dt} &= R_a^{\star},
\end{align}
with $\mathbf F^{\star}$ the drag, $Q_c^{\star}$ and $Q_r^{\star}$ convective and radiative
sources, $Q_p^{\star}$ the evaporative sink and $R_a^{\star}$ the accretion rate.

The melt fraction $X_{\mathrm{melted}}^{\star}$ is a smooth regularized step of
$T^{\star}$ around the fusion temperature, so that the latent heat of fusion $L_f$ enters
an effective particle heat capacity
\begin{equation}
  C_p^{\star} = C_{p,\mathrm s}
  + X_{\mathrm{melted}}^{\star}\,(C_{p,\ell} - C_{p,\mathrm s})
  + L_{f}\,\frac{d X_{\mathrm{melted}}^{\star}}{dT^{\star}}.
\end{equation}
with $C_{p,\mathrm s}$ and $C_{p,\ell}$ the solid- and liquid-phase specific heats of the
particle.

Drag is $\mathbf F^{\star} = (m^{\star}/\tau^{\star})(\mathbf u - \mathbf v^{\star})$, with the
response time $\tau^{\star}$ from a Mach-dependent drag law: Haider--Levenspiel below slip Mach
$0.3$ and Loth~\cite{loth2008compressibility}, which retains compressibility and rarefaction
effects, above it.
Radiation is $Q_r^{\star} = \epsilon_p\,\sigma\,A^{\star}(T_{\mathrm{rad}}^4 - T^{\star 4})$,
with $\epsilon_p$ the particle emissivity, $\sigma$ the Stefan--Boltzmann constant, $A^{\star}$
the particle surface area and $T_{\mathrm{rad}}$ the effective radiation temperature.
Once liquid, evaporation follows Stefan--Fuchs~\cite{sazhin2014heating},
\begin{equation}
  R_a^{\star} = -\pi d^{\star}\,\frac{\lambda_r}{C_{p,r}}\,
  \frac{\mathrm{Sh}}{\mathrm{Le}}\,\log(1 + B_M),
\end{equation}
where the Sherwood number follows from the Nusselt number by the heat--mass-transfer analogy,
$\mathrm{Sh}=\mathrm{Nu}\,\mathrm{Le}^{-1/3}$, and the Nusselt number uses the
Richter--Nikrityuk correlation for non-spherical particles~\cite{richter2012drag}, which
reduces to the Ranz--Marshall form for a sphere ($S^{\star}=1$). Here $d^{\star}$ is the
particle diameter, $\lambda_r$ and $C_{p,r}$ the thermal conductivity and specific heat of the
vapor--gas film, $\mathrm{Le}$ the Lewis number, $B_M$ the Spalding mass-transfer number, and
$\mathrm{Re}_r$ and $\mathrm{Pr}$ the film Reynolds and Prandtl numbers; the film properties
are evaluated at an Eckert temperature~\cite{eckert1956engineering}. The heat sink is
$Q_p^{\star} = -R_a^{\star} L_v$, with $L_v$ the latent heat of vaporization. Liquid droplets
break up when the gas Weber number
$\mathrm{We}_g = \rho\,U_{\mathrm{rel}}^2\,r^{\star}/\sigma^{\star}$ is large (with
$U_{\mathrm{rel}}$ the gas--particle slip velocity, $r^{\star}$ the droplet radius and
$\sigma^{\star}$ its surface tension), using a
KH/RT model~\cite{beale1999modeling,reitz1987modeling,patterson1998modeling}.
Particle--shock interaction in the diamond train is left for future work.

\subsection{Boundary conditions}
\label{sec:bc}
The remaining pieces of $\partial\Omega$ are the axis $\Gamma_{\mathrm{axis}}$, the gun-exit
inlet $\Gamma_{\mathrm{in}}$, the nozzle outlet $\Gamma_{\mathrm{out}}$, the outer wall
$\Gamma_{\mathrm{w}}\subset\partial\Omega_s$, and the coil curve
$\Gamma_{\mathrm{coil}}\subset\partial\Omega_s$.

\paragraph{Axial symmetry.}
The twin is posed in the $(r,z)$ half-plane. On $\Gamma_{\mathrm{axis}}$ ($r=0$) we impose
symmetry: $\mathbf u\cdot\mathbf n=0$, $\mathbf q\cdot\mathbf n=0$.

\paragraph{Inlet.}
On $\Gamma_{\mathrm{in}}$ the mass flow $\dot m$ is prescribed. The net power $P_{\mathrm{in}}$
(torch power minus cooling losses) then fixes the incoming enthalpy flux through the surface
integrals
\begin{align}
  \dot m
  &= \int_{\Gamma_{\mathrm{in}}} \rho\,\mathbf u\cdot\mathbf n\,\mathrm{d}A, \\[1pt]
  P_{\mathrm{in}}
  &= \int_{\Gamma_{\mathrm{in}}} \rho\,\mathbf u\cdot\mathbf n\,
     \bigl[H(T)+\tfrac12|\mathbf u|^{2}\bigr]\,\mathrm{d}A.
\end{align}
The inlet velocity $\mathbf u$ and temperature $T$ are prescribed implicitly, uniform over
$\Gamma_{\mathrm{in}}$, so that both integral constraints above are verified. Here $H(T)$ is
the specific enthalpy of the LTE Ar/H$_2$ mixture: Cantera~\cite{cantera} tabulates it versus
temperature at the prescribed composition, from the equilibrium mass fractions as
$H=\sum_i Y_i h_i(T)$. $P_{\mathrm{in}}$ is inferred in Section~\ref{sec:AI} rather than
measured at the gun exit.

\paragraph{Outlet.}
$\Gamma_{\mathrm{out}}$ is an outlet where the pressure $p=75$\,mbar is imposed where the flow
leaves subsonically and left free where it leaves supersonically, selected automatically by
the local Mach number.

\paragraph{Walls.}
No-slip $\mathbf u=\mathbf 0$ is imposed on $\Gamma_{\mathrm{fs}}$. Heat conduction in the
solid and conjugate matching read
\begin{align}
  \rho_s C_{p,s}\,\partial_t T + \nabla\!\cdot\mathbf q &= 0
    \quad\text{in }\Omega_s,
    \quad \mathbf q = -\lambda_s\nabla T, \\[1pt]
  T\big|_{\Omega_f}=T\big|_{\Omega_s},
  \quad
  \mathbf q_f\!\cdot\!\mathbf n
  &=\mathbf q_s\!\cdot\!\mathbf n
    \quad\text{on }\Gamma_{\mathrm{fs}}.
\end{align}
The solid is a 316L stainless-steel vessel, cooled by the copper coil below
(\emph{Coil cooling}); at $300$\,K, $\rho_s=8238\,\mathrm{kg\,m^{-3}}$,
$C_{p,s}=468\,\mathrm{J\,kg^{-1}K^{-1}}$ and
$\lambda_s=13.4\,\mathrm{W\,m^{-1}K^{-1}}$~\cite{incropera2006}.

\paragraph{Coil cooling.}
The helical copper coil is reduced to a 1-D coolant problem on $\Gamma_{\mathrm{coil}}$,
parametrized by arc length $s$,
\begin{equation}
  \partial_t T_{\mathrm{coil}}
  + v_{\mathrm{w,proj}}\,\partial_s T_{\mathrm{coil}}
  = \omega_c\,\bigl(T\big|_{\Gamma_{\mathrm{coil}}} - T_{\mathrm{coil}}\bigr).
\end{equation}
$T_{\mathrm{coil}}(s,t)$ is the coolant temperature and $T|_{\Gamma_{\mathrm{coil}}}$ the wall
temperature restricted to the coil curve; $v_{\mathrm{w,proj}}$ is the coolant speed projected
onto that curve. The wall flux is
$Q_b = -h_{\mathrm{wall}}(T - T_{\mathrm{coil}})$. The exchange rate
$\omega_c = h_{\mathrm{tot}}\,\pi d_i/(\rho_w c_{p,w} A_i)$ is distinct from the SST
dissipation $\omega$, with $d_i$ and $A_i$ the inner diameter and cross-section of the tube
and $(\rho_w,c_{p,w})$ the coolant density and specific heat. Resistances add in series,
$1/h_{\mathrm{tot}} = 1/h_{\mathrm{conv}} + 1/h_{\mathrm{contact}} + 1/h_{\mathrm{Cu}}$.
The convective (Dittus--Boelter) and copper-conduction terms come from the literature,
whereas the coil-to-wall contact conductance $h_{\mathrm{contact}}$ is by far the most
uncertain: it depends on how well the coil is pressed against the vessel. We therefore split
it off and infer it from experiment (Section~\ref{sec:calib}); the recovered value is
physically meaningful, quantifying the quality of the coil-to-solid thermal contact.

\paragraph{Particles.}
The feedstock enters through the radial powder-injection port (Figure~\ref{fig:reactor})
following the measured PSD and sphericity of Section~\ref{sec:particles}. A particle that
reaches $\Gamma_{\mathrm{out}}$ is counted as collected; one that reaches a wall is held there
by COMSOL's \emph{Freeze} condition, and the accumulated frozen mass gives the local wall
deposition rate (\si{mm\,min^{-1}}).

% ====================================================================
%  DISCRETIZATION
% ====================================================================
\section{Discretization}
\label{sec:numerics}

The high-Mach interface is a stabilized FEM: equal-order P1--P1 velocity--pressure
elements with consistent streamline and crosswind diffusion~\cite{comsol2024cfd}. Time
integration of the Eulerian twin is implicit (BDF), so the time step follows the convective
residence time of the chamber rather than the acoustic CFL set by the speed of sound.

The domain is a free triangular mesh of about $65{,}000$ elements
(Figure~\ref{fig:reactor}), refined in the nozzle throat where the flow chokes and at the
walls with eight boundary-layer rows so that $\Delta_w^{+}<1$, as required by the low-Re SST
closure. That wall and throat resolution is the mesh design criterion and the standard
requirement of the turbulence model; it is the resolution at which the calibration
observables (chamber pressure, coil heat) are obtained. The twin is a time-dependent study
matched to the experimental series, with inlet power ramped from a cold start.

Particles are traced in a separate study with the generalized-$\alpha$
method~\cite{chung1993alpha,jansen2000alpha}: an implicit second-order scheme that
dissipates unresolved high frequencies while preserving the slow thermal ODE of each
particle. They are injected at the last gas-flow time step, into the preheated background; the
\emph{Specify mass flow rate} option sets about $1{,}000$ representative particles to the true
feed rate, and the study is advanced until every particle has left through the outlet or frozen
on a wall. The remaining mismatch with the experiments then sits in uncertain PDE
parameters, which Section~\ref{sec:AI} identifies from the rig.

% ====================================================================
%  DATA ASSIMILATION
% ====================================================================

\section{Data-Driven Calibration, Discovery and Surrogates}
\label{sec:AI}

The predictive accuracy of COMSOL is limited by three factors: uncertain parameters, missing physics, and computational cost. We address them with AI methods and models that (i)~calibrate the solver against physical experiments, (ii)~identify and quantify the missing physics, and (iii)~provide fast neural-operator surrogates. COMSOL then serves both as a solver and as a data generator.

\subsection{Derivative-free parameter calibration}
\label{sec:calib}
The parameters of the PDE can be inferred through experimental recordings by solving an inverse problem. All uncertain parameters are collected in a vector $\boldsymbol{\theta}\in\mathbb{R}^d$ and inferred by minimizing the least-squares objective function $\mathcal{J}$, which measures the difference
between simulated outputs $\mathcal{G}(\boldsymbol{\theta})$ and sensor data
$\mathbf{y}_{\mathrm{obs}}$,
\begin{equation}
    \mathcal{J}(\boldsymbol{\theta})
    = \frac{1}{2}
      \bigl\| \mathbf{y}_{\mathrm{obs}} - \mathcal{G}(\boldsymbol{\theta}) \bigr\|^2.
\end{equation}
Gradient-based methods need $\partial\mathcal{G}/\partial\boldsymbol{\theta}$, which commercial
solvers often do not expose. We therefore treat COMSOL as a black-box operator.
We leverage Unscented Kalman Inversion (UKI)~\cite{singh2024inverse}, which propagates $2d+1$ sigma points
and approximates the parameter--observable cross-covariance without an explicit Hessian. Furthermore, we include regularization inspired on the Levenberg--Marquardt~\cite{nocedal2006numerical}, which adds adaptive regularization, balancing robustness
and convergence when each COMSOL evaluation is expensive. This combination identifies a stable second-order optimization approach, which is feasible due to the limited number of parameters $d$ (roughly a dozen parameters) and the ability to instantiate multiple COMSOL runs in parallel.

\subsection{Learned source terms}\label{sec:AI_source}
After calibration, we aim to identify and quantify unmodeled physical phenomena. Hence, we augment the
COMSOL time step with a neural source $\mathcal{S}_{\mathrm{NN}}(\mathbf{u};\boldsymbol{\phi})$,
\begin{equation}
    \mathbf{u}^{n+1}
    = \mathbf{u}^n
      + \Delta t \, \mathbf{F}(\mathbf{u}^n)
      + \Delta t \, \mathcal{S}_{\mathrm{NN}}(\mathbf{u}^n; \boldsymbol{\phi}),
\end{equation}
where $\mathbf{u}^{n}$ is the discrete solution state at time step $n$, $\Delta t$ the time
step, $\mathbf{F}$ the deterministic COMSOL update (the nominal discretized right-hand side)
and $\boldsymbol{\phi}$ the trainable network weights. The source is
trained against sensor trajectories. Notably, a trained neural source term is beneficial in data assimilation as it can account for noise in sensor measurements. Furthermore, neural source terms act as implicit regularizers in the time stepping of numerical solvers \cite{koehler2026neural}.

The learned source is then analyzed with
SINDy~\cite{brunton2016discovering} against a library of candidate terms (diffusion,
kinetics, radiative $T^4$, advection), recovering a sparse expression of the missing
physics and guiding later COMSOL refinements.

\subsection{Neural surrogates}
\label{sec:surrogates}
Neural operators \cite{kovachki2023neural} trained on COMSOL results can be used to map inlet and geometry parameters to the resulting physical fields in milliseconds. Motivated by the observed quasi-monotonic topology of reactor fields over various geometries, we can expect a trained operator to
generalizes across simulation conditions without exhaustive sampling. This enables real-time application of the solver, and represents the reduced-order layer used for monitoring and process optimization.

% ====================================================================
%  RESULTS
% ====================================================================
\section{Simulation Results}
\label{sec:results}
All the simulations presented use a single AMD EPYC 9454P (96 cores, 251\,GB RAM). The Eulerian gas/solid transient
solve of about $4.1\times10^{5}$ degrees of freedom takes $\sim$20\,min, and the subsequent
particle-tracing study about $10$\,min. At this problem size the parallel efficiency of a
single solve saturates near $24$ cores; total throughput is therefore maximized by running
several solves side by side rather than one large solve, so calibration sweeps use four
concurrent $24$-core jobs instead of a single $96$-core job.

\subsection{Fluid fields}
\label{sec:fluidfields}

\begin{figure}[H]
  \centering
  \includegraphics[width=0.84\linewidth]{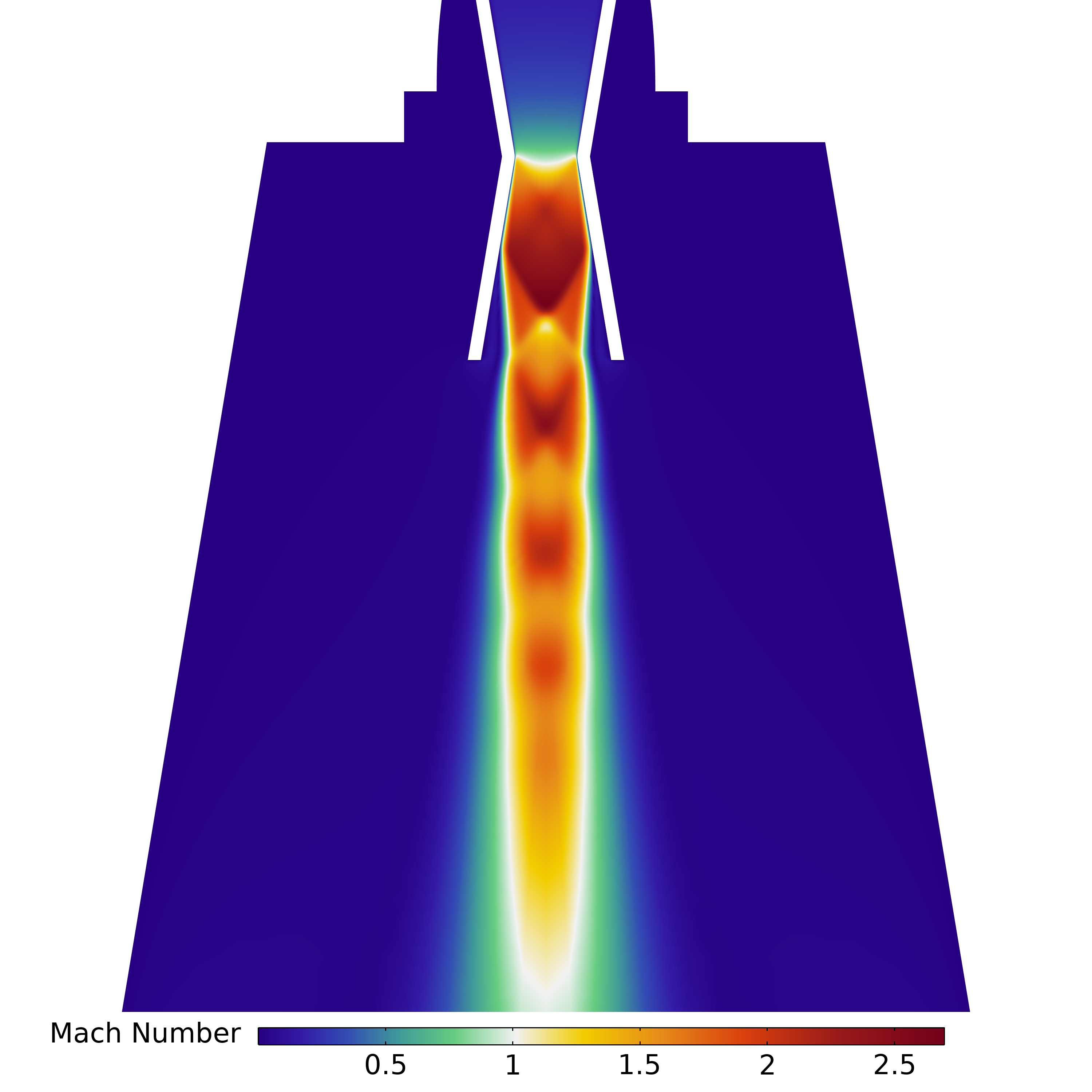}
  \caption{Mach-number field in the converging--diverging nozzle and supersonic jet
  (Ma $\approx2$--$3$), with the shock-diamond train in the divergent section.}
  \label{fig:mach}
\end{figure}

Figure~\ref{fig:mach} shows the corresponding
Mach-number field: the flow accelerates through the throat and expands into a supersonic jet
(Ma $\approx 2$--$3$) with a clear shock-diamond train.

\begin{figure}[H]
  \centering
  \includegraphics[width=\linewidth]{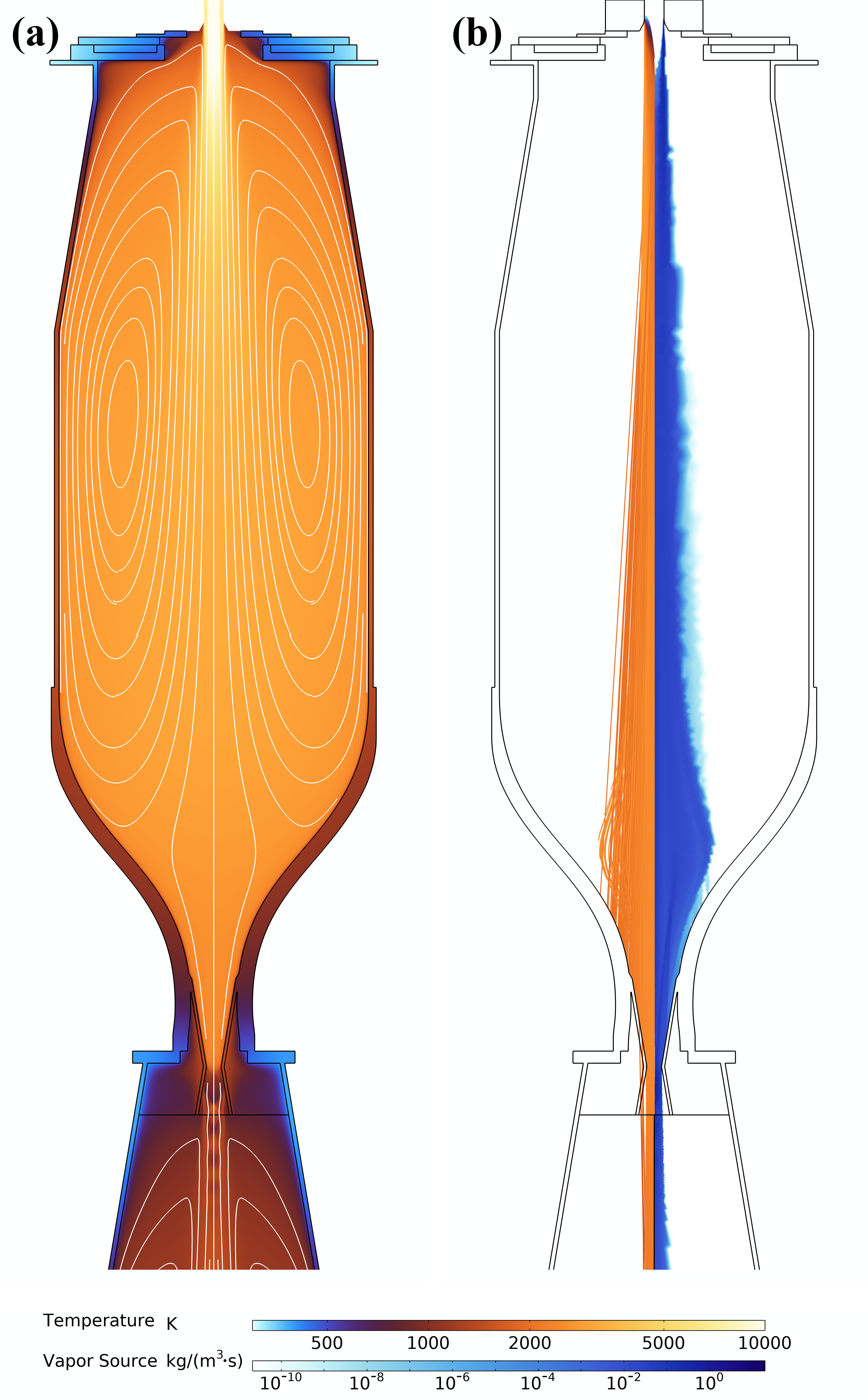}
  \caption{COMSOL twin at a representative operating point
  (Ar\,100/H\textsubscript{2}\,2.5\,NLPM, $P_{\mathrm{in}}\approx22$\,kW). (a)~gas temperature
  field with flow streamlines. (b)~Ti64 particle trajectories with the vapor source they
  release into the gas. The temperature colorbar is shared by the gas field (a) and the
  particle trajectories (b).}
  \label{fig:fields}
\end{figure}

Figure~\ref{fig:fields}(a) shows the temperature field at a representative Ar/H\textsubscript{2}
operating point: the
arc-heated jet enters at the top at more than $12\,000$\,K, fills the chamber, and cools as it
is drawn through the converging--diverging nozzle, while the water-cooled walls stay near
ambient. This steep core-to-wall gradient makes local probe readings unreliable and
motivates matching global integrals rather than local probes
(Section~\ref{sec:discussion}). This is the background into which the particles are
released.

\subsection{Particle evolution and product powder}
\label{sec:powder}
The model produces, for every injected particle, the full time history of temperature,
melting, and evaporation as it travels through the reactor. Figure~\ref{fig:fields}(b) shows a
representative Ti64 run. The particle trajectories, colored by temperature (left half), cross
the liquidus almost immediately after injection into the hot core. The mass the particles lose
to evaporation is accumulated into a volumetric vapor source (right half), which is diluted
into the main gas at low molar fraction.

The shock-diamond train of Figure~\ref{fig:mach} sets the slip that drives Mach-dependent
drag and KH/RT breakup. A dedicated high-slip-Mach model for particle--shock interaction in
that train remains future work (Section~\ref{sec:particles}).

\begin{figure}[H]
  \centering
  \includegraphics[width=\linewidth]{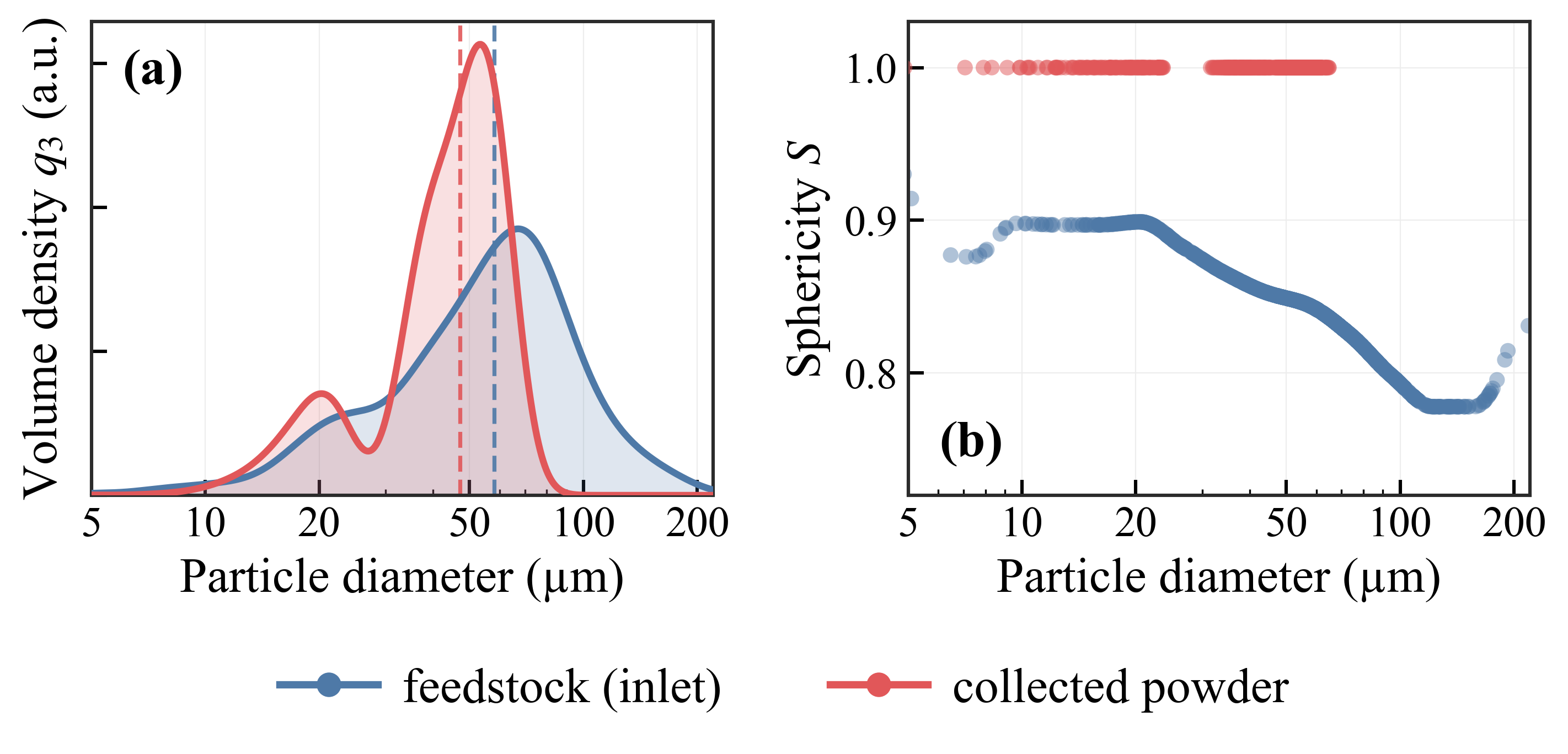}
  \caption{Predicted product powder, feedstock versus collected. (a)~volume PSD $q_3$
  (dashed lines mark $D_{50}$). (b)~sphericity versus diameter.}
  \label{fig:powder}
\end{figure}

At the collection plane the model returns a diameter and a sphericity for every particle,
the metrics the process is judged on. Distributions are weighted by each particle's release
frequency and mass. Figure~\ref{fig:powder}(a) compares feedstock and collected volume
PSDs: evaporation depletes the coarse ($>100\,\mu$m) tail and shifts the median from
$D_{50}\approx58$ to $47\,\mu$m. Figure~\ref{fig:powder}(b) shows the shape change: irregular
feedstock ($S_0\approx0.78$--$0.93$) melts and leaves as spheres ($S=1$). Calibration of
the particle-side model (evaporation, breakup, loss) against collected powder is left for
future work (Section~\ref{sec:calib}).

% ====================================================================
%  DISCUSSION
% ====================================================================
\section{Discussion}
\label{sec:discussion}

\subsection{Calibration against experiment}
Since the optimization results on all the uncertainties of the PDEs represent a trade secret of GREEN14, we showcase the calibration results of the method explained in Section~\ref{sec:calib} on two parameters: the power actually delivered to the gas,
$P_{\mathrm{in}}$, and the coil-to-wall contact conductance $h_{\mathrm{contact}}$. We optimize those by matching two measured integrals: the
chamber pressure $p$, and the heat taken by the coil
$Q_{\mathrm{coil}}=\dot m_w c_p\Delta T$. Local probes sit in a field that drops from
$\sim$15\,000\,K over a few centimeters, so they depend too strongly on placement; $p$ and
$Q_{\mathrm{coil}}$ are global and nearly independent ($p$ mainly constrains
$P_{\mathrm{in}}$, $Q_{\mathrm{coil}}$ mainly constrains $h_{\mathrm{contact}}$). The
subsonic chamber is almost isobaric, so $p$ is essentially the chamber stagnation pressure.
Since the nozzle is choked, the chamber pressure and the throat gas temperature are tightly
coupled~\cite{anderson2003modern}, so the measured $p$ is a reliable proxy for the gas
temperature.

\begin{figure}[H]
  \centering
  \includegraphics[width=\linewidth]{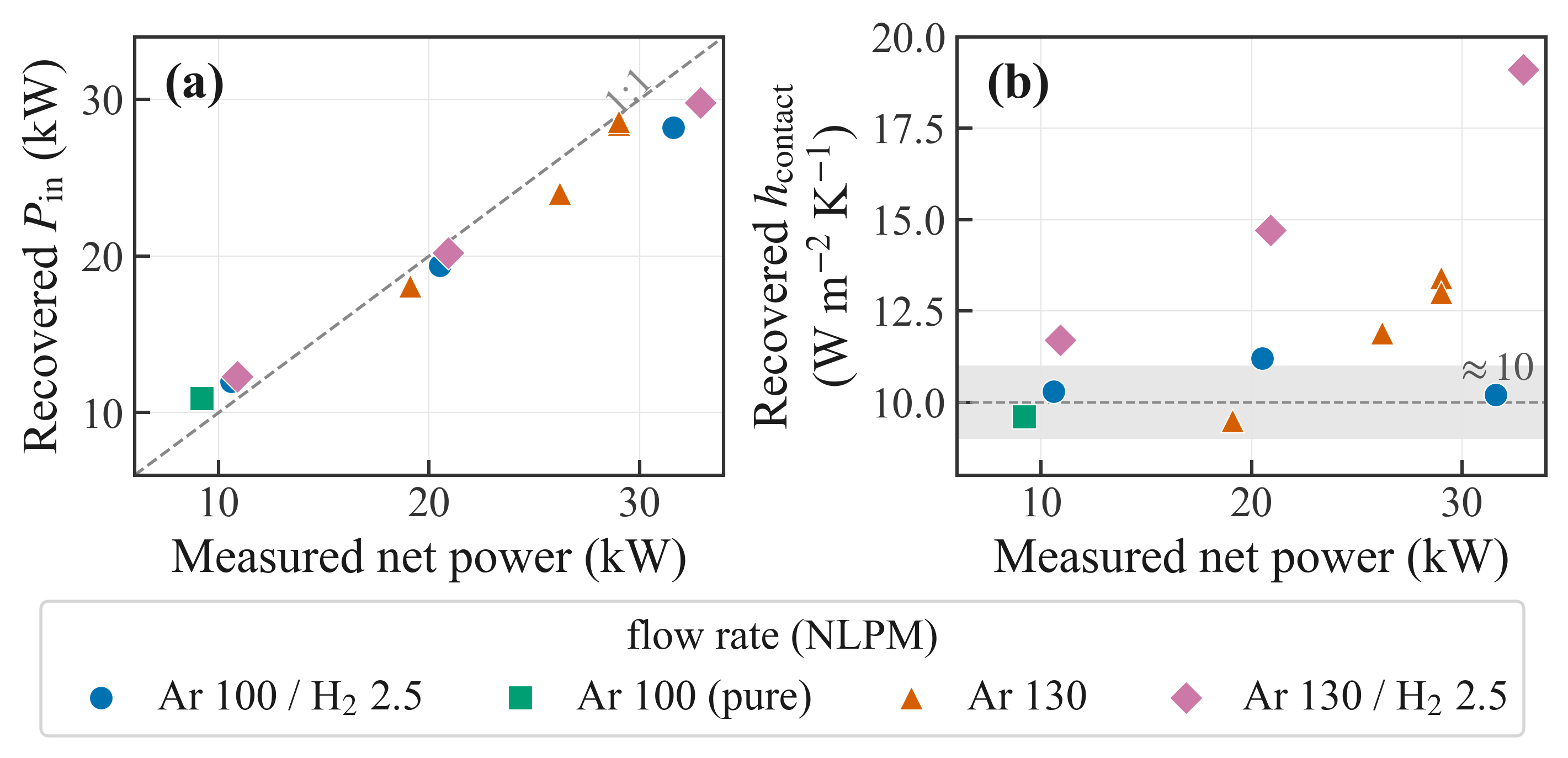}
  \caption{Calibration across all hot-gas experiments, colored by gas mixture. (a)~recovered
  $P_{\mathrm{in}}$ versus measured net power (dashed = equality). (b)~recovered
  $h_{\mathrm{contact}}$ versus power (shaded band at $10\,\mathrm{W\,m^{-2}K^{-1}}$).}
  \label{fig:calib}
\end{figure}

Applied across three gas mixtures and a threefold range of power (Figure~\ref{fig:calib};
mixtures labeled by Ar/H$_2$ NLPM, e.g.\ Ar\,100/H\textsubscript{2}\,2.5), the recovered
$P_{\mathrm{in}}$ tracks the measured net power even though power was never a fit target,
and $h_{\mathrm{contact}}$ stays near $10\,\mathrm{W\,m^{-2}K^{-1}}$, in agreement with
gun-off cooldown calorimetry. Over 11 experiments with hot gas the chamber pressure is matched to $0.10\%$
on average ($0.30\%$ worst) and the coil power to $0.42\%$ ($1.13\%$ worst). This
self-consistency, rather than any single matched test, is what validates the twin. The one
mixture where $h_{\mathrm{contact}}$ drifts with power flags a composition-dependent
discrepancy. The same strategy was applied to selected thermocouples and to
$T_{\mathrm{coil}}$.

\subsection{Predictive use of the simulations}
Once calibrated against the rig (Section~\ref{sec:calib}), the twin reproduces the operating
reactor closely enough to be used predictively, not only to reconstruct a finished run. Because
it resolves the full Eulerian state, it exposes quantities the rig cannot instrument and turns
them into operating margins: (i)~the local thermal and condensation state in the throat
(Figures~\ref{fig:fields},~\ref{fig:mach}), together with the wall deposition rate computed
from the frozen particles (Section~\ref{sec:bc}), signals incipient nozzle clogging
(a common failure in high-temperature metallurgy) before the throat area is restricted;
(ii)~the pressure
field gives the chamber-to-exit drop $\Delta p$, hence the expansion efficiency of a given
operating point; (iii)~the conjugate solution locates wall and coil hot spots and, by comparing the predicted
wall temperature against the melting point of the 316L vessel ($\approx1400\,^{\circ}$C),
turns them into a quantitative margin to hardware damage while there is still time to react.

\subsection{Physics-oriented identification and live use}
The ML methods of Section~\ref{sec:AI} act on the physics model itself: a small set of
uncertain coefficients $\boldsymbol{\theta}$ is inferred from the experiments
(Section~\ref{sec:calib}), and missing mechanisms are recovered as a sparse source in the
governing equations (Section~\ref{sec:AI_source}). The twin that is marched forward remains
the COMSOL PDE, with identified parameters and terms.

Once that model is aligned with the rig, it can be queried live and used for design.
First, it informs process engineers in real time during a run: reactor state, hot spots,
and whether the nozzle and chamber remain in a safe window. Second, it is used to
optimize the system, first in software and then in hardware: geometry and operating
parameters toward a process objective (a target PSD, energy extracted in the quench, and
similar). GREEN14 has already gone through two hardware optimization iterations driven by
this loop and is currently working toward surrogate-based reactor controls for
self-supervised experiments.

% ====================================================================
%  CONCLUSIONS
% ====================================================================
\section{Conclusions}
We developed an axisymmetric multiphysics COMSOL model of the GREEN14 argon--hydrogen
plasma reactor, combining compressible RANS, conjugate heat transfer and Lagrangian
phase-change particles. The PDEs are the standard, robust COMSOL closures, representative
of the physical system (about $20$\,min for the Eulerian solve and $10$\,min for the
particles) and intended for process-engineer support and scale-up.

The remaining gap with the rig is closed by inferring a small set of uncertain PDE
coefficients from experiment, rather than by training a network on a library of simulated
fields. Across 11 hot-gas experiments the twin matches chamber pressure to $0.10\%$ on
average ($0.30\%$ worst) and coil power to $0.42\%$ ($1.13\%$ worst). Once aligned, the
fields can be used predictively (nozzle clogging, $\Delta p$ efficiency, wall hot spots)
and have already driven two hardware iterations at GREEN14.

Missing source terms are recovered as sparse analytic expressions in the governing
equations (Section~\ref{sec:AI_source}); a neural-operator layer
(Section~\ref{sec:surrogates}) is only a millisecond wrapper around that identified twin.
In live use, sensor readings feed a model on the process engineer's computer; offline the
same model optimizes software and then hardware. The results shown here are for Ti64; the
framework extends to other powders and to other industrial plasma processes.

{\footnotesize
\bibliographystyle{IEEEtran}
\bibliography{references}}

@article{qiu2023triple,
  author  = {Qiu, Jier and Yu, Deping and Xiao, Yu and Fan, Ying and Chen, Yiwen and Li, Dingjun},
  title   = {A novel triple-cathode plasma torch with hot-wall nozzle for {YSZ} spherical
             thin-walled hollow-shell powder preparation},
  journal = {Ceramics International},
  volume  = {49},
  number  = {16},
  pages   = {27551--27566},
  year    = {2023},
  doi     = {10.1016/j.ceramint.2023.06.030}
}

@article{hao2021rfti64,
  author  = {Hao, Zhenhua and Hou, Xuchu and Zhang, Qinglei and Zhu, Xingying and
             Zhou, Fa and Shu, Yongchun and He, Jilin},
  title   = {Preparation of spherical {Ti-6Al-4V} powder by {RF} induction plasma
             spheroidization combined with spray granulation},
  journal = {Powder Technology},
  volume  = {387},
  pages   = {88--94},
  year    = {2021},
  doi     = {10.1016/j.powtec.2021.04.021}
}

@article{liu2026plasma,
  author  = {Liu, Kun and Yu, Deping and Liu, Jinwei and Yao, Yimeng and He, Juntao and
             Liu, Dawei and Jia, Shuaihang},
  title   = {Constrained flow-field regulation in in-flight droplet plasma atomization for
             stable production of spherical metal powders},
  journal = {Powder Technology},
  volume  = {485},
  pages   = {122988},
  year    = {2026},
  doi     = {10.1016/j.powtec.2026.122988}
}

@inproceedings{Menter2003SSTHeatTransfer,
  author    = {F. Menter and J.C. Ferreira and T. Esch and B. Konno},
  title     = {The {SST} Turbulence Model with Improved Wall Treatment for Heat Transfer Predictions in Gas Turbines},
  booktitle = {Proceedings of the International Gas Turbine Congress (IGTC)},
  pages     = {IGTC2003-TS-059},
  year      = {2003},
  address   = {Tokyo, Japan}
}

@manual{comsol2024particle,
  title   = {Particle Tracing Module User's Guide, ``Droplet Breakup Theory''},
  author  = {{COMSOL Multiphysics}},
  edition = {Version 6.4},
  year    = {2024}
}

@article{haider1989drag,
  title     = {Drag coefficient and terminal velocity of spherical and nonspherical particles},
  author    = {Haider, A. and Levenspiel, O.},
  journal   = {Powder Technology},
  volume    = {58},
  number    = {1},
  pages     = {63--70},
  year      = {1989},
  publisher = {Elsevier}
}

@article{richter2012drag,
  title     = {Drag forces and heat transfer coefficients for spherical, cuboidal and
               ellipsoidal particles in cross flow at sub-critical {R}eynolds numbers},
  author    = {Richter, Andreas and Nikrityuk, Petr A.},
  journal   = {International Journal of Heat and Mass Transfer},
  volume    = {55},
  number    = {4},
  pages     = {1343--1354},
  year      = {2012},
  publisher = {Elsevier}
}

@article{loth2008compressibility,
  title     = {Compressibility and rarefaction effects on drag of a spherical particle},
  author    = {Loth, E.},
  journal   = {AIAA Journal},
  volume    = {46},
  number    = {9},
  pages     = {2219--2228},
  year      = {2008},
  publisher = {American Institute of Aeronautics and Astronautics},
  doi       = {10.2514/1.28943}
}

@article{eckert1956engineering,
  title   = {Engineering relations for heat transfer and friction in high-velocity laminar and turbulent boundary-layer flow over surfaces with constant pressure and temperature},
  author  = {Eckert, E. R. G.},
  journal = {Transactions of the ASME},
  volume  = {78},
  pages   = {1273--1283},
  year    = {1956}
}

@incollection{sazhin2014heating,
  title     = {Heating and evaporation of multicomponent droplets},
  author    = {Sazhin, Sergei},
  booktitle = {Droplets and Sprays},
  pages     = {143--178},
  year      = {2014},
  publisher = {Springer}
}

@article{reitz1987modeling,
  title   = {Modeling atomization processes in high-pressure vaporizing sprays},
  author  = {Reitz, Rolf D.},
  journal = {Atomisation and Spray Technology},
  volume  = {3},
  pages   = {309--337},
  year    = {1987}
}

@article{patterson1998modeling,
  title     = {Modeling the effects of fuel spray characteristics on diesel engine combustion and emission},
  author    = {Patterson, Mark A. and Reitz, Rolf D.},
  journal   = {SAE Transactions},
  pages     = {27--43},
  year      = {1998},
  publisher = {JSTOR}
}

@article{beale1999modeling,
  title     = {Modeling spray atomization with the {Kelvin--Helmholtz}/{Rayleigh--Taylor} hybrid model},
  author    = {Beale, Jennifer C. and Reitz, Rolf D.},
  journal   = {Atomization and Sprays},
  volume    = {9},
  number    = {6},
  pages     = {623--650},
  year      = {1999},
  publisher = {Begel House Inc.}
}

@manual{comsol2024cfd,
  title   = {{CFD Module User's Guide}, ``The High Mach Number Flow Interfaces'' and ``Theory for the Turbulent Flow Interfaces''},
  author  = {{COMSOL}},
  edition = {Version 6.4},
  year    = {2024}
}

@article{Menter1994SST,
  title     = {Two-equation eddy-viscosity turbulence models for engineering applications},
  author    = {Menter, F. R.},
  journal   = {AIAA Journal},
  volume    = {32},
  number    = {8},
  pages     = {1598--1605},
  year      = {1994},
  publisher = {American Institute of Aeronautics and Astronautics},
  doi       = {10.2514/3.12149}
}

@misc{cantera,
  author       = {David G. Goodwin and Harry K. Moffat and Ingmar Schoegl and
                  Raymond L. Speth and Bryan W. Weber},
  title        = {{Cantera}: An Object-oriented Software Toolkit for Chemical
                  Kinetics, Thermodynamics, and Transport Processes},
  year         = {2023},
  note         = {Version 3.0.0},
  doi          = {10.5281/zenodo.8137090},
  howpublished = {\url{https://www.cantera.org}}
}

@article{wargnier2020transport,
  author  = {Wargnier, Q. and Alvarez Laguna, A. and Scoggins, J. B. and
             Mansour, N. N. and Massot, M. and Magin, T.},
  title   = {Consistent transport properties in multicomponent two-temperature
             magnetized plasmas: Application to the {Sun} atmosphere},
  journal = {Astronomy \& Astrophysics},
  volume  = {635},
  pages   = {A87},
  year    = {2020},
  doi     = {10.1051/0004-6361/201834686}
}

@article{wargnier2022collision,
  author  = {Wargnier, Q. M. and Mart{\'i}nez-Sykora, J. and Hansteen, V. H.
             and De Pontieu, B.},
  title   = {Detailed Description of the Collision Frequency in the Solar Atmosphere},
  journal = {The Astrophysical Journal},
  volume  = {933},
  number  = {2},
  pages   = {205},
  year    = {2022},
  doi     = {10.3847/1538-4357/ac6e62}
}

@article{cressault2010nec,
  author  = {Cressault, Y. and Rouffet, M. E. and Gleizes, A. and Meillot, E.},
  title   = {Net emission of {Ar}--{H}$_2$--{He} thermal plasmas at atmospheric pressure},
  journal = {Journal of Physics D: Applied Physics},
  volume  = {43},
  number  = {33},
  pages   = {335204},
  year    = {2010},
  doi     = {10.1088/0022-3727/43/33/335204}
}

@book{incropera2006,
  author    = {Incropera, Frank P. and DeWitt, David P. and Bergman, Theodore L.
               and Lavine, Adrienne S.},
  title     = {Fundamentals of Heat and Mass Transfer},
  edition   = {6th},
  year      = {2006},
  publisher = {John Wiley \& Sons},
  address   = {Hoboken, NJ}
}

@book{anderson2003modern,
  author    = {Anderson, John D.},
  title     = {Modern Compressible Flow: With Historical Perspective},
  edition   = {3rd},
  year      = {2003},
  publisher = {McGraw-Hill},
  address   = {New York}
}

@article{singh2024inverse,
  title={Inverse unscented Kalman filter},
  author={Singh, Himali and Mishra, Kumar Vijay and Chattopadhyay, Arpan},
  journal={IEEE Transactions on Signal Processing},
  volume={72},
  pages={2692--2709},
  year={2024},
  publisher={IEEE}
}

@book{nocedal2006numerical,
  title={Numerical optimization},
  author={Nocedal, Jorge and Wright, Stephen J},
  year={2006},
  publisher={Springer}
}

@book{sapunov2024deep,
  title={Deep learning with JAX},
  author={Sapunov, Grigory},
  year={2024},
  publisher={Simon and Schuster}
}

@article{kovachki2023neural,
  title={Neural operator: Learning maps between function spaces with applications to pdes},
  author={Kovachki, Nikola and Li, Zongyi and Liu, Burigede and Azizzadenesheli, Kamyar and Bhattacharya, Kaushik and Stuart, Andrew and Anandkumar, Anima},
  journal={Journal of Machine Learning Research},
  volume={24},
  number={89},
  pages={1--97},
  year={2023}
}

@article{wadell1935,
  author  = {Wadell, Hakon},
  title   = {Volume, Shape, and Roundness of Quartz Particles},
  journal = {The Journal of Geology},
  volume  = {43},
  number  = {3},
  pages   = {250--280},
  year    = {1935}
}

@article{chung1993alpha,
  author  = {Chung, J. and Hulbert, G. M.},
  title   = {A Time Integration Algorithm for Structural Dynamics With Improved
             Numerical Dissipation: The Generalized-$\alpha$ Method},
  journal = {Journal of Applied Mechanics},
  volume  = {60},
  number  = {2},
  pages   = {371--375},
  year    = {1993},
  doi     = {10.1115/1.2900803}
}

@article{jansen2000alpha,
  author  = {Jansen, Kenneth E. and Whiting, Christian H. and Hulbert, Gregory M.},
  title   = {A generalized-$\alpha$ method for integrating the filtered
             {Navier--Stokes} equations with a stabilized finite element method},
  journal = {Computer Methods in Applied Mechanics and Engineering},
  volume  = {190},
  number  = {3--4},
  pages   = {305--319},
  year    = {2000},
  doi     = {10.1016/S0045-7825(00)00203-6}
}

@article{brunton2016discovering,
  title={Discovering governing equations from data by sparse identification of nonlinear dynamical systems},
  author={Brunton, Steven L and Proctor, Joshua L and Kutz, J Nathan},
  journal={Proceedings of the national academy of sciences},
  volume={113},
  number={15},
  pages={3932--3937},
  year={2016},
  publisher={National Academy of Sciences}
}

@article{koehler2026neural,
  title={Neural emulator superiority: When machine learning for PDEs surpasses its training data},
  author={Koehler, Felix and Thuerey, Nils},
  journal={Advances in Neural Information Processing Systems},
  volume={38},
  pages={165660--165702},
  year={2026}
}

\section*{Acknowledgements}
The authors thank the GREEN14 team for the pilot-reactor measurements used to validate the model.

\end{document}